\documentclass[aps,reprint,prl,twocolumn,superscriptaddress]{revtex4}

\usepackage{epsfig,dsfont,amssymb,amsmath,amsthm,amsfonts,amsbsy,mathrsfs}
\usepackage{graphicx}
\usepackage{color}
\usepackage[dvipsnames]{xcolor}
\usepackage{bm}
\usepackage{multirow}
\usepackage{comment}
\usepackage{hyperref}
\usepackage{natbib}
\usepackage{breqn}

\newcommand {\comm}[1]{{\color{green} #1}}

\begin{document}

\title{Phase Coexistence and Edge Currents in the Chiral Lennard-Jones Fluid}


\author{Claudio B. Caporusso}
\email{claudio.caporusso@ba.infn.it}
\affiliation{Dipartimento Interateneo di Fisica\char`,{} Universit\`a degli Studi di Bari and INFN Sezione di Bari\char`,{} via Amendola 173 Bari 70126 Italy}
\author{Giuseppe Gonnella}
\affiliation{Dipartimento Interateneo di Fisica\char`,{} Universit\`a degli Studi di Bari and INFN Sezione di Bari\char`,{} via Amendola 173 Bari 70126 Italy}
\author{Demian Levis}
\affiliation{Departament de Física de la Mat\`eria Condensada\char`,{} Universitat de Barcelona C. Martí Franqu\`es\char`,{} 1 08028 Barcelona Spain.}
\affiliation{UBICS University of Barcelona Institute of Complex Systems Mart\'i i Franqu\`es\char`,{} 1 E08028 Barcelona Spain.}

\date{\today}

\begin{abstract} 
We study a model chiral fluid in two dimensions composed of Brownian disks interacting via a Lennard-Jones potential and a non-conservative transverse force, mimicking colloids spinning  at a  rate $\omega$. 
The system exhibits a phase separation between a chiral liquid and a dilute gas phase that can be characterized using a thermodynamic framework. We compute the equations of state and show that the surface tension controls interface corrections to the coexisting pressure predicted from the equal-area construction.
Transverse forces increase surface tension and generate edge currents at the liquid-gas interface. The analysis of these currents shows that the rotational viscosity introduced in chiral hydrodynamics is consistent with microscopic bulk mechanical measurements. Chirality can also break the solid phase, giving rise to a dense fluid made of rotating hexatic patches. Our work paves the way for the development of the statistical mechanics of chiral particles assemblies.  

\end{abstract}

\maketitle


Chiral fluids composed of interacting spinning particles break  parity and time-reversal symmetry,  
 defying our fundamental description of soft materials \cite{liebchen2022chiral}. 
 Salient examples include suspensions of magnetic colloids driven by an external rotating field \cite{snezhko2016complex, soni2019odd, massana2021arrested, joshi2022extension, bililign2022motile}, 
 and chiral grains put into motion by an inner motor or a vibrated plate \cite{scholz2018rotating,workamp2018symmetry,yang2020robust, lopez2022chirality,vega2022diffusive}. In all cases a common feature arises: transverse pairwise forces. 
In spinning colloids, they result from the advection of the flow field generated by their rotation \cite{lenz2003membranes,massana2021arrested}, while in dry systems,  it is the friction between colliding grains that lead to them \cite{tsai2005chiral, han2021fluctuating}. 
As spinning colloids carry a permanent magnetic moment they self-assemble into clusters that eventually coarsen, and the resulting drops sustain steady edge currents at their surface  \cite{soni2019odd, massana2021arrested}.  
The dynamic properties of such chiral liquid interface have been described within hydrodynamic models, allowing to measure odd transport coefficients that capture the breakdown of parity symmetry in the constitutive equations \cite{banerjee2017odd, soni2019odd}, akin odd viscosity in quantum Hall systems \cite{avron1995viscosity, avron1998odd}. The possibility to control edge states in soft matter  has boosted the interest on these systems, both from experiments and theory \cite{fruchart2023odd}. 
%
Localized currents have been also observed in dry systems at the interface between two  species with opposite chirality \cite{nguyen2014emergent, scholz2018rotating} or at the boundaries of a confining wall \cite{van2016spatiotemporal, yang2020robust,yashunsky2022chiral}, raising the question about the impact of transverse forces on the properties of matter. Despite recent progress, there is still much to  understand about the nature of phase transitions in chiral particle systems, as well as the minimal ingredients needed to sustain such edge currents and their impact on the large-scale behaviour.  

\begin{figure}[ht!]
    \centering
    \includegraphics[width=0.95\linewidth]{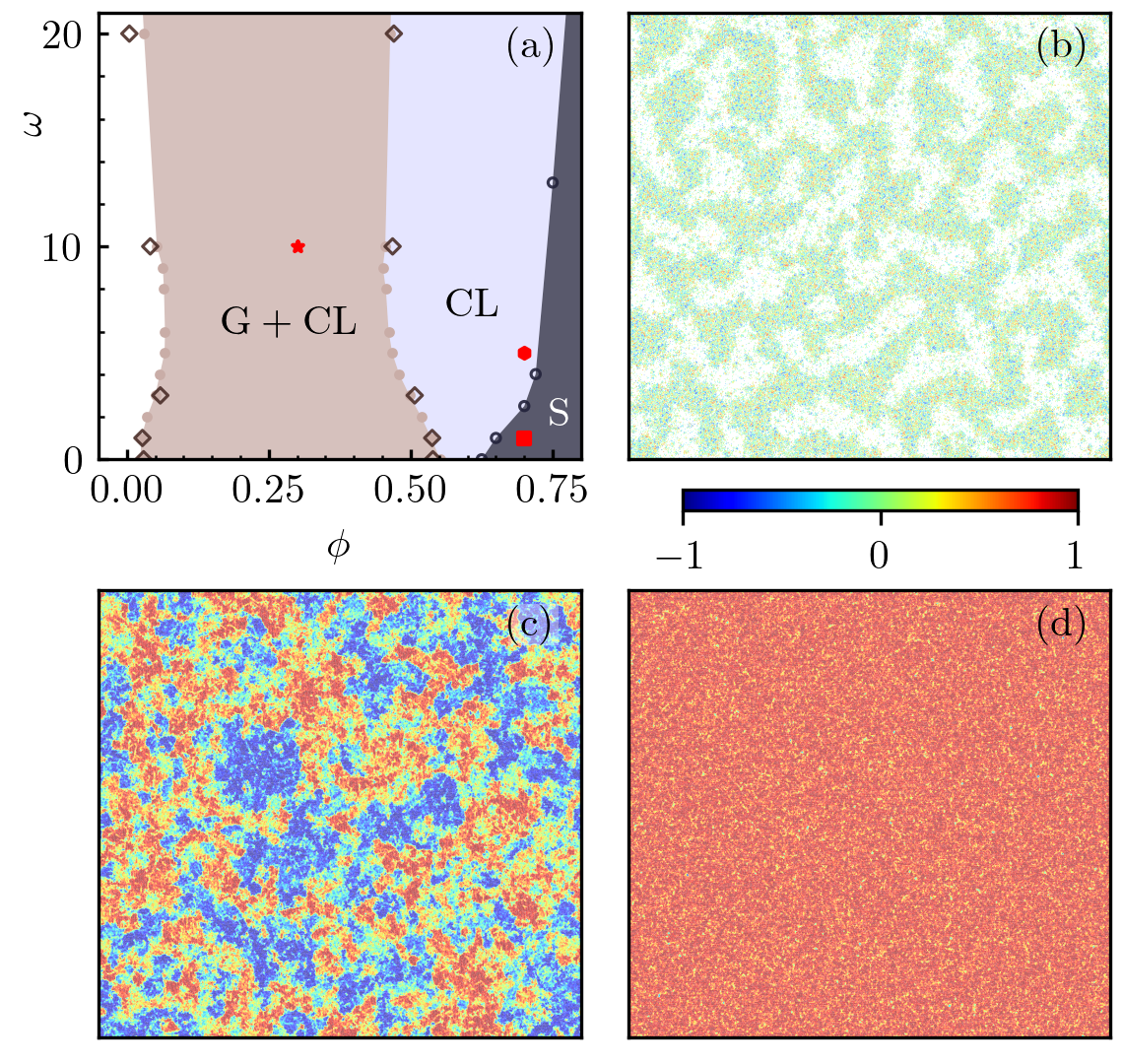}
    \caption{ 
    (a) Phase diagram at $T=0.47$ comprising a gas  (white), chiral liquid (blue) and solid  (grey) phases, and a coexistence region (brown). Filled symbols are obtained from density distributions while the empty ones from the pressure.  Black circles come from the analysis of the hexatic order parameter.
     (b)-(d) Snapshots of the system corresponding to 
      $(\phi,\,\omega) =(0.3, 10)$ (b), $(0.7, 
 5)$ (c) and $(0.7, 1)$ (d). Particles are colored according to the projection of the local hexatic order $\psi_{6}(r_i)$ on the global mean orientation.  
}
    \label{fig:coexistence}
\end{figure}

To  address such question, we study a  chiral extension of the  Lennard-Jones model in two dimensions (2D) \cite{barker1981phase, rovere1990gas, smit1991vapor, PICACIAMARRA-2D}, with pairwise, non-conservative, transverse forces. Although dipolar interactions are responsible for the clustering of  magnetic colloids \cite{jager2012nonequilibrium, jaeger2013dynamics, snezhko2016complex, massana2021arrested}, here we  consider a paradigmatic reference model incorporating attractive and excluded volume interactions to decipher the impact of self-spinning on general grounds. 
Using extensive numerical simulations we establish its phase behaviour, identifying a coexistence region between a dilute gas and a \emph{chiral liquid} (G+CL), followed by a (homogeneous) chiral liquid  (CL) and finally a \emph{chiral solid} phase (S), see Fig. \ref{fig:coexistence}. Despite being out-of-equilibrium, we show that  phase separation in this system can be well characterized using equilibrium concepts. However, the resulting chiral liquid shows steady edge currents at its interface, akin to the ones observed in colloidal spinner suspensions. From the analysis of such currents one can estimate the rotational viscosity, that can also be computed in bulk from the microscopic stress tensor. 
Upon self-spinning, we find that the Lennard-Jones solid breaks into a mosaic of dynamic hexatic patches with a finite characteristic size, as observed in experiments \cite{bililign2022motile}.



We consider $N=512^2$ disks located at $\bm{r}_{i}$ moving in a {$V = L_x \times L_y$ } box with periodic boundary conditions (PBC) and obeying the Langevin equation
\begin{dmath}
    m \ddot{\bm{r}}_i + \Gamma \dot{\bm{r}}_i = \sum_{j(\neq i)} [\bm{f}_{ij}-\bm{\nabla}_i U(r_{ij})]  +  \sqrt{2 \Gamma k_B T} \bm{\nu}_i,
    \label{eq:model}
\end{dmath}
where $m$ is the mass of the particles, $\Gamma$  the damping coefficient, $T$  the temperature and $\bm{\nu}_i$  a Gaussian white noise with zero mean and unit variance. 
The disks interact via the  Lennard-Jones potential
$U(\bm{r}) = \epsilon \left[ (\sigma_d/r)^{12} - (\sigma_d / r)^6 \right] + \epsilon$, truncated at $r_c = 5\sigma_d$. 
Inspired by colloidal spinners \cite{massana2021arrested}, the disks experience a transverse force {$\bm{f}_{ij} = \zeta \frac{\omega}{\omega_0} \hat{z} \times \bm{r}_{ij} / r_{ij}^3$} only if $r_{ij}<r_c$, where $\hat{z}$ is the normal vector to the plane where particles move, {$\tau=1/\omega_0 = \sqrt{m\sigma_d^2/\epsilon}$ fixes the time unit  and $\zeta = m \sigma_d^2 / \tau^3$}.  The parameter $\omega$ thus quantifies the chirality and the departure from equilibrium.  
We fix $\Gamma = 100$ and $m = \sigma_d = \epsilon = 1$, providing the units of mass, length and energy, respectively, and
allowing us to adimensionalize all the observables. Within this parameters range, the dynamics can be considered overdamped.  We integrate Eq.~\ref{eq:model}  using a velocity-Verlet algorithm  \cite{LAMMPS} over a broad range of surface fractions $\phi = \pi N/4V$ and chiralities $\omega$ at fixed $k_BT = 0.35$ and $0.47$ (details in   \cite{SM}).  


To investigate the phase behavior of our model we let the system relax from  initial configurations where particles are arranged in a hexagonal lattice (with the spacing fixed by $\phi$).
Fig.~\ref{fig:coexistence}(a) shows the resulting phase diagram in the $(\phi,\omega)$ plane at a fixed  $k_BT=0.47$ (see \cite{SM} for the one at $k_BT = 0.35$). In the  $\omega \rightarrow 0$  limit we recover the equilibrium  2D Lennard-Jones phase diagram  \cite{PICACIAMARRA-2D}.  
As the density is increased from the  homogeneous gas, 
a phase separation, inherited from the equilibrium liquid-gas demixing, occurs for all the values of $\omega$ explored. We identify a coexistence region (brown area in Fig. \ref{fig:coexistence} (a)) in which dense liquid droplets form and coarsen in a gas background (see Fig. \ref{fig:coexistence} (b) and \cite{SM}).
The coexisting densities are not significantly altered by $\omega$. 
Beyond, a chiral liquid phase exists for all the $\omega$ values explored, characterized by collective flows and short-range hexatic order, see Fig.~\ref{fig:coexistence}(c). As the density increases the hexatic correlation length grows and eventually diverges, signalling the presence of a chiral solid phase, see Fig.~\ref{fig:coexistence}(c). 
 The liquid-solid transition shifts towards higher $\phi$ with $\omega$. 

\begin{figure}
    \centering
    \includegraphics[width=\linewidth]{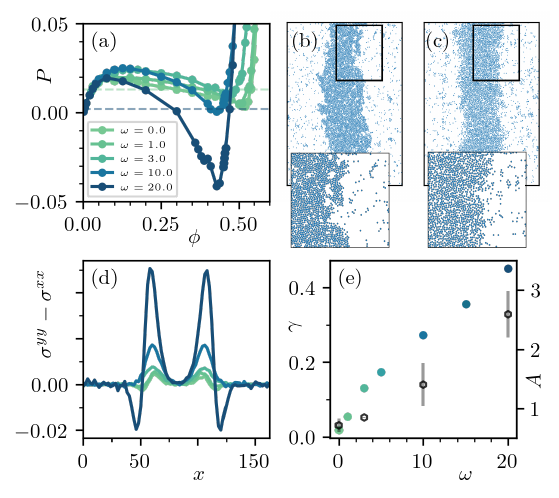}
    \caption{
    (a) Equation of state for different values of $\omega$ at $k_BT=0.47$. The dotted lines correspond to the equal-area construction applied to $\omega=0$, $20$ curves. (b-c) Slab configurations at (b) $\omega = 0$ and (c) $\omega = 20$ with a detailed view. (d) Corresponding profiles of the difference $\sigma^{xx} - \sigma^{yy}$ and resulting (e) surface tension $\gamma$ (left y-scale), and area $A$ of the  pressure loop ($\times 10^3$) (right y-scale).
    }
    \label{fig:EOS}
\end{figure}

Bringing thermodynamic tools to this new context, we first estimate the coexistence boundary (or binodals) from the density-dependence of the mechanical pressure $ P=- {\mbox{Tr}(\sigma)}/{2} $, given by the trace of the  Irving-Kirkwood 
stress tensor \cite{irving1950statistical}:
\begin{equation}\label{eq:irving}
   {\sigma^{ab} = -  \frac{1}{V}\sum_i m v^a_i v^b_i - \frac{1}{2V} \sum_{i,j\neq i} r^a_{ij}  F_{ij}^b,\, }
\end{equation}
where $\bm{F}_{ij}=\bm{f}_{ij}-\bm{\nabla}_i U(r_{ij})$
and $a=x,y$ a cartesian component. As $\bm{f}_{ij}$ is transverse, it does not bring any explicit contribution to the pressure (but it does  implicitly via the statistics of collisions) and $P(\phi)$  defines an \emph{equation of state},  
shown in Fig.~\ref{fig:EOS}(a)  for different values of $\omega$.  By construction, the ideal gas law remains unchanged in the dilute limit. 
Then, the pressure shows a double loop structure characteristic of phase coexistence.
In equilibrium, these Mayer-Wood loops are an interface effect in finite size systems, as phase separation brings a free energy excess $\Delta \mathcal{F}=\gamma \ell $, where $\gamma$ is the interface tension and $\ell$ the length of the interface. At moderate $\phi$, a liquid drop becomes asymptotically circular, 
while at higher densities, the liquid drop closes onto itself via PBC, resulting in a slab with    two flat interfaces, see Fig.\ref{fig:EOS} (c,d). 
As for the pressure loops, 
 the  binodals 
 can be obtained through the Maxwell, or equal-area, construction. Although such construction cannot be readily applied for $\omega\neq 0$, we use it by extension (as done to identify  the coexistence region in systems of active particles \cite{solon2015pressure, levis2017active, cugliandolo2017phase, digregorio2018full, solon2018generalized}). The  coexistence pressures we obtain for $\omega=0$ and $20$ are shown in Fig. \ref{fig:EOS} (a). As shown in Fig~\ref{fig:coexistence}(a), we find a very good agreement between  the location of the binodals  obtained from  the equal-area construction and the analysis of the density distribution (shown in \cite{SM}). 

Interestingly, the area of the Mayer-Wood loops  increases with $\omega$.  Pushing the thermodynamic mindset further, this would mean that the free energy excess due to the interface is larger in the chiral system. As we are out-of-equilibrium, we can't compute the interface tension from the free energy excess, but should rely on a purely statistical mechanical definition \cite{kirkwood1949statistical, evans1979nature}. It can be defined as the difference between the normal $\sigma^{xx}$ and transverse $\sigma^{yy}$ pressure across a flat interface 
\begin{dmath}
    \gamma = -\frac{1}{2}\int_{0}^{L_x} dx \left[ \sigma^{xx}(x) - \sigma^{yy}(x) \right].
    \label{eq:s_tension}
\end{dmath} 
To compute $\gamma$ we prepare the system under conditions for which the interfaces are flat (on average, along $y$), removing the Laplace pressure contribution. 
Such slab configurations  Fig. \ref{fig:EOS} (b, c) show that the chiral liquid interface is smoother than the equilibrium one. This visually confirms the interpretation that, upon self-spinning, the interface tension increases. To quantify this claim, we compute the stress profile along the $x$-direction normal to the interface, averaging over the tangential $y$-component. 
Fig.~\ref{fig:EOS}(d) displays the difference between the two across the slab, showing a similar behaviour as in equilibrium Lennard-Jones  \cite{trokhymchuk1999computer, mejia2005phase} and active particles systems 
\cite{bialke_negative, paliwal2017non, hermann2019phase, lauersdorf2021phase}. As expected, the normal and tangential pressure (reported in \cite{SM}) are identical and constant in the bulk phases. The tangential one exhibits then a moderate drop followed by a fast increase across the interface, signalling that it is first smaller then larger than the normal pressure as we cross the interface from the gas. Such behaviour, present in equilibrium,  becomes more pronounced as $\omega$ increases.
The explanation relies on the establishment of edge currents (see discussion below), that enhance particle collisions at the interface in the tangential direction. 
  As reported in Fig.~\ref{fig:EOS}(e), the excess tangential pressure results in a growth of $\gamma$ with $\omega$, that roughly follows the same trend as the area of the pressure loops, supporting the thermodynamic  description of phase coexistence developed so far.  

\begin{figure}
    \vspace*{0.25cm}
    \centering
    \includegraphics[width=\linewidth]{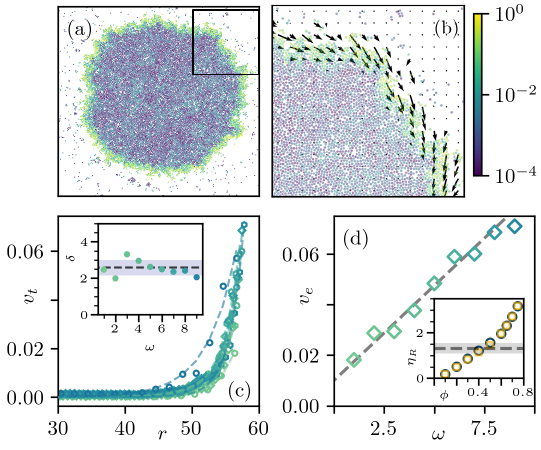}
    \caption{ 
    (a) A chiral drop at  $T = 0.35$ and $\omega = 5.0$.  
    Particles are colored according to their velocity (normalized by the fastest particle one, in log scale). 
    (b) Detailed view of the configuration showing  the 
    velocity field. 
    (c) Tangential velocity as a function of the distance from  the droplet center for $\omega=1...9$. The exponential fits (discontinuous lines) allow to extract the values of $\delta$  shown in the inset, with the dotted line representing the average value $\delta = 2.57 \pm 0.42$.
    (d) Dependence of the edge velocity $v_e$ 
    with $\omega$, showing a linear growth. 
     Inset: 
    $\eta_R(\phi)$, from the  computation of $\sigma^{xy}$ 
    for $\omega=3.0$ (yellow) and $\omega=10.0$ (blue) together with the estimation from the edge currents  $\eta_R = 1.32 \pm 0.22$ (dotted line).}
    \label{fig:edge_currents}
\end{figure}

As mentioned earlier, transverse forces generate edge currents. 
As shown in Fig. \ref{fig:edge_currents}(a,b) a localized flow of particles establishes in our system, for any $\omega$, at the liquid-gas interface. In the hydrodynamic description, chirality introduces a \emph{rotational viscosity} $\eta_R$ that 
 can be estimated from the analysis of the edge currents \cite{soni2019odd}. 
Indeed,   hydrodynamic models predict a localized flow at the surface of a chiral liquid drop, that decays exponentially with a characteristic penetration length $\delta = \sqrt{(\eta_0 + \eta_R)/\mu}$, where $\eta_0$ and $\eta_R$ are the shear and rotational viscosity, respectively, and $\mu$ the friction coefficient  (given here by the damping coefficient $\Gamma$, see \cite{SM}). 
Moreover, the edge velocity $v_e$, i.e. the tangent velocity of particles $v_t$ right at the chiral drop interface, can also be written in terms of $\eta_R$, as $v_{e} \approx 2\omega\delta\frac{\eta_R}{\eta_0+\eta_R}$. Overall, one can estimate the rotational viscosity from the analysis of the particles' velocity profile  within a chiral drop, as
$ \eta_R = {\Gamma \delta v_e}/({2 \omega})$.

To measure  edge currents, we let a spherical drop relax to its steady state at a fixed $\omega$. We then measure the velocity field  $\bm{v}(t) = \delta \bm{r}(t)/\delta t$ from particle displacements $\delta \bm{r}(t) = \bm{r}(t+\delta t) - \bm{r}(t)$ over a  time interval $\delta t = 10^3\; \tau$. 
%
\footnote{We checked indeed that the time interval $\delta t$ for computing the displacement field was larger than the typical relaxation times of the particles inside the droplet} %
The system then reaches a steady edge current, see Fig.~\ref{fig:edge_currents}(a-b) and movie M2 \cite{SM}. We then coarse-grain the tangential component of the velocity over square cells, and spherically averaged it, obtaining a one-dimensional profile $v_{t}(r)$ that depends only on the distance from the center of mass of the droplet.  
Fig.~\ref{fig:edge_currents}(c) shows these profiles: $v_{t}(r)$ is almost zero across the bulk of the droplet, and increases exponentially towards the edge. We then fit our data 
and find that, in agreement with hydrodynamics and experiments \cite{soni2019odd}, $\delta$ appears to be independent of $\omega$ while $v_e$ grows linearly, see Fig.~\ref{fig:edge_currents}(d). Combining  $\delta$ and $v_e$ we extract a rotational viscosity $\eta_R\approx 1.32$, reported in Fig.~\ref{fig:edge_currents}(d). 
One can also put the hydrodynamic description aside and provide a  measurement of $\eta_R$ from the microscopic dynamics, provided by the  expression of the stress tensor in Eq.~\ref{eq:irving}:  $\eta_R=\sigma^{xy}/2\omega$. The results are reported in Fig.~\ref{fig:edge_currents}(d) for different $\phi$ and $\omega$. They show that $\eta_R$ grows with $\phi$ but has little dependence on $\omega$. For $\phi = 0.6$ - slightly above the  density of the dense drop in this case - the value of $\eta_R$ extracted in the homogeneous chiral fluid from the microscopic stress tensor, $\eta_R\approx 1.9$,  is  reasonably close the one estimated from the edge currents in the coexistence regime. 
As in the homogeneous chiral liquid phase, one cannot rely on the hydrodynamic description of surface flows to estimate $\eta_R$. The microscopic expression of $\eta_R$ we used is most useful, as it provides a systematic route to  compute it in the absence of distinct interfaces \cite{soni2019odd}.      

\begin{figure}[h!]
    \centering
    \includegraphics[width=0.97\linewidth]{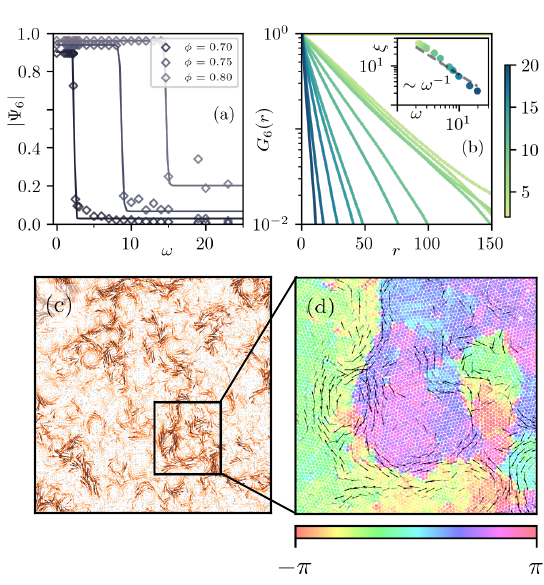}
    \caption{ 
    (a) Global hexatic order $|\Psi_6|$  as a function of $\omega$ at fixed $k_BT=0.35$,  
    for three different densities. The solid line is a hyperbolic tangent  fit. 
    (b) Exponential decay of $G_6(r)$ for systems at $\phi=0.70$ and different values of $\omega$, shown in the colormap. Inset: corresponding correlation lengths $\xi$. (c) Displacement field $\delta \bm{r}$ in a chiral liquid configuration ($k_BT=0.35$, $\phi=0.70$, $\omega = 3.0$), (darker  arrows indicate larger displacements). (d) Detailed view showing the local bond orientation $\text{Arg}(\psi_6)$ together with  particle displacements. }
    \label{fig:patterns}
\end{figure}    

In equilibrium, the system is solid for densities higher than a threshold $\phi_s$ ($\approx 0.6$ at $k_BT=0.35$ and $\approx 0.62$ at $k_BT=0.47$)
 (Fig. \ref{fig:coexistence} (d)). 
By turning on $\omega$, we introduce extra stresses that can eventually melt the solid, breaking its long-range orientational order and resulting in a  chiral liquid state composed of rotating hexatic patches, see Fig. \ref{fig:patterns} (c,d) and movie M3 \cite{SM}. 
 To quantify it, we use the hexatic order parameter $\psi_6(\bm{r}_j)=n_j^{-1}\sum_{k\in \partial_j} e^{i6\theta_{jk}}$, where $\theta_{jk}$ is the angle formed by the segment connecting the center of the jth disk with a nearest neighbor (out of $n_j$) and the x-axis. As shown in Fig. \ref{fig:patterns} (a), the global hexatic order parameter $|\Psi_6| =  \sum_{i=1}^N \left| \psi_{6i} \right|/N$ exhibits  sharp drop at a critical value $\omega_c$, that shifts to higher values with $\phi$.
We determine $\omega_c$ with a hyperbolic tangent fit  and report its values in the phase diagram Fig~\ref{fig:coexistence} (a) as delimiting a solid from a chiral liquid phase. We compute the correlation function $G_6(r) = \langle \psi_6(r) \psi_6(0) \rangle$ to quantify the typical size of the hexatic domains observed \cite{negro2022hydrodynamic, caporusso2020motility}. As shown in Fig.~\ref{fig:patterns}(b), $G_6$ decays exponentially, defining a characteristic length scale $\xi$ that scales as $\sim \omega^{-1}$. Such decay of a typical length associated with a spatial pattern induced by chirality has also been found in chiral active matter models \cite{liebchen2017collective, sese2022microscopic}, planar rotors \cite{rouzaire2021defect} and experiments of chiral colloids \cite{massana2021arrested,bililign2022motile}. 

The dynamics of the dense chiral liquid is highly heterogeneous, as particles close to the hexatic domain walls are  mobile while the ones within remain roughly frozen. 
The snapshot, Fig.~\ref{fig:patterns}(c), together with its detailed view, Fig.~\ref{fig:patterns}(d), showing both the displacement field and the local orientation of the hexatic domains, $\text{Arg}(\psi_{6,i})$, show that particle flow is localized around the boundaries of the domains, while inside each patch particles are frozen. 
 These domains are a consequence of the transverse interaction, as in equilibrium one would get global hexatic order in this regime, 
and  are responsible for the fluidization of the system. The domains can slide over each other, resulting in a 'swirling motion' akin to what is observed in experiments \cite{bililign2022motile}. This mechanism gives rise to a rich and complex dynamics, whose precise characterisation calls for further investigation.

Introducing transverse forces to an otherwise standard Lennard-Jones system, 
is enough to generate edge currents and, at large densities,  a mosaic of rotating domains with a characteristic length, as observed in experiments \cite{soni2019odd, massana2021arrested, bililign2022motile}. The breakdown of parity symmetry 
 leads  to a rotational viscosity, a transport coefficient that can be measured from the microscopic stress tensor, providing results compatible with the predictions of hydrodynamic models. This gives hope for the development of systematic derivations of Irving-Kirwood kind of formulas relating the transport coefficients appearing in the constitutive equations of chiral systems and the microscopic particle variables \cite{fruleux2016mesoscopic}. The impact of chirality on the liquid-gas transition can be well rationalized using the thermodynamics of phase coexistence: chirality increases the liquid-gas interface tension, giving rise to larger Mayer-Wood loops in the equations of state, that allows to locate the binodals, applying the equal-area construction. Overall, our work provides a simple framework to investigate how chirality affects the liquid and solid phases of matter from a statistical mechanics viewpoint.

\paragraph*{Acknowledgments.}

We thank for the access to the HPC resources ReCaS in Bari and MareNostrum4 at the BSC.
C.C. and G.G. acknowledge MIUR project PRIN/2020 PFCXPE ``Response, control and learning: building new manipulation strategies in living and engineered active matter". 

\nocite{caporusso2020motility}

\bibliography{chiral_abbrv}

\end{document}




\renewcommand{\thefigure}{S\arabic{figure}}
\renewcommand{\theequation}{S\arabic{equation}}
\setcounter{figure}{0}

\title{Supplemental Material of ``Phase Coexistence and Edge Currents in the Chiral Lennard-Jones Fluid" }
\maketitle

In this Supplemental Material, we provide additional details and results that support the  findings presented in the main text. 


\section{Numerical details}\label{sec:model}

We performed molecular dynamics simulations in the NVT ensemble. We used a velocity Verlet algorithm with a Langevin-type thermostat to integrate the equation of motion given by Eq. (1) in the main text.  We started from hexagonally ordered configurations, with the distance between particles adjusted according to the global density, and let them relax under the influence of the interactions until they reached a steady-state, starting from which we measured all the quantities reported in the paper. We employed as  units the mass $m$, the energy $\epsilon$ and the length $\sigma_d$ of the particles, which we made dimensionless and set equal to $1$. All the other units are expressed in terms of these base units, e.g., the time unit is $\tau = \sqrt{m\sigma_d^2/\epsilon}$. We used the parameters $\gamma = 100$, and $k_B T = 0.35$ or $k_B T= 0.47$. The timestep of the simulation was chosen to be $dt = 0.01 \tau$, which ensured numerical stability for the choice of parameters described before. The size of the system was fixed by the choices of the global packing fraction $\phi$ and of the number of particles, which we set to $N = 512^2$. The size ratio between the linear dimensions was set to $L_x / L_y = \sqrt{3}/2$. We modified the open-source software LAMMPS to implement the integration of our model. A typical simulation of $\approx 10^4\; \tau$ required 72 hours using 96 CPUs. We found that for large enough values of $\omega$, typically $\omega > 20$, the thermostat could not efficiently dissipate the energy generated by the interaction and this translates into heating up of the system, and for this reason, we limited the range of the parameter below this value. 

\section{Movies}


We provide three movies that illustrate the phase behaviour and collective dynamics of chiral particles in different geometries. The movies are available as supplementary material and can be accessed online at \url{https://www.dropbox.com/sh/l1tkekz2w9ktkld/AACiZAo4Io3havaC_fQNQp0Ta?dl=0}. The movies show the following:

\begin{enumerate}
\item The phase separation  process from an initial homogeneous configuration for a system with $N=512^2$, $\phi = 0.20$ and $\omega = 10.0$ with $k_B T=0.47$.
\item The edge current in a phase-separated chiral drop. Parameters used: $N=128^2$, $k_B T=0.35$ and $\omega = 5.0$. The particles are colored according to the displacement field magnitude. 
\item The breaking of the solid phase and the melting transition induced by increasing  $\omega$ in a system with $N=512^2$ particles with $k_B T=0.35$ and $\omega = 3.0$. The movie displays the particle positions and a color map of the local bond orientational parameter of each particle, $\text{Arg}[{\psi_{6}(\boldsymbol{r}_i)}]$.

\end{enumerate}

\newpage
\FloatBarrier


\section{Phase diagram at $k_B T=0.35$}

\begin{figure}[!htbp]
    \centering
    \includegraphics[width=0.4\linewidth]{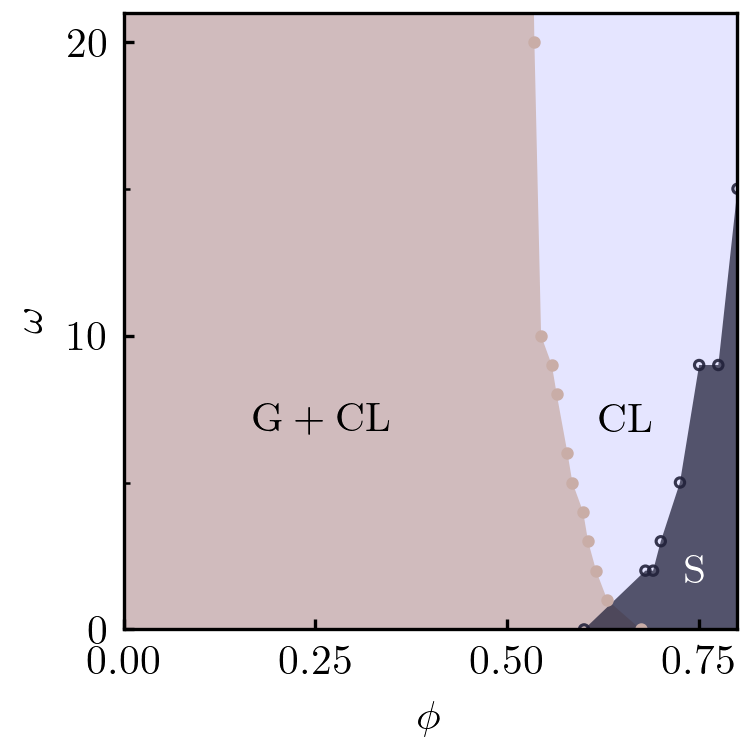}
     \caption{The phase diagram of the model at $k_BT = 0.35$, showing the coexistence region between the chiral liquid and gas phases (brown shaded area), the chiral liquid phase (blue shaded area), and the solid phase (dark shaded area). The brown dots indicate the points where the peak of the local density was measured, 
     and the black circles indicate the ones obtained from the behaviour of the global hexatic order parameter $|\Psi_6|$.}    \label{SM:phasediagram}
\end{figure}

We report here the phase diagram of the model at $k_B T = 0.35$, a lower temperature than the one in the main text ($k_B T = 0.47$). Figure \ref{SM:phasediagram} shows the phase diagram in the $(\phi, \omega)$ plane, obtained by measuring the peaks of the local density distribution (see the relative section later in the text)
and the hexatic order parameter as a function of density and chiral interaction strength $\omega$. We observe that the coexistence region between the chiral liquid and gas phase expands as the temperature decreases, up to reaching $\phi = 0$ for the lower branch and overlapping the solid transition for the higher density branch at small $\omega$. 


\section{Structure factor}

\begin{figure}[!htbp]
    \centering
    \includegraphics[width=0.75\linewidth]{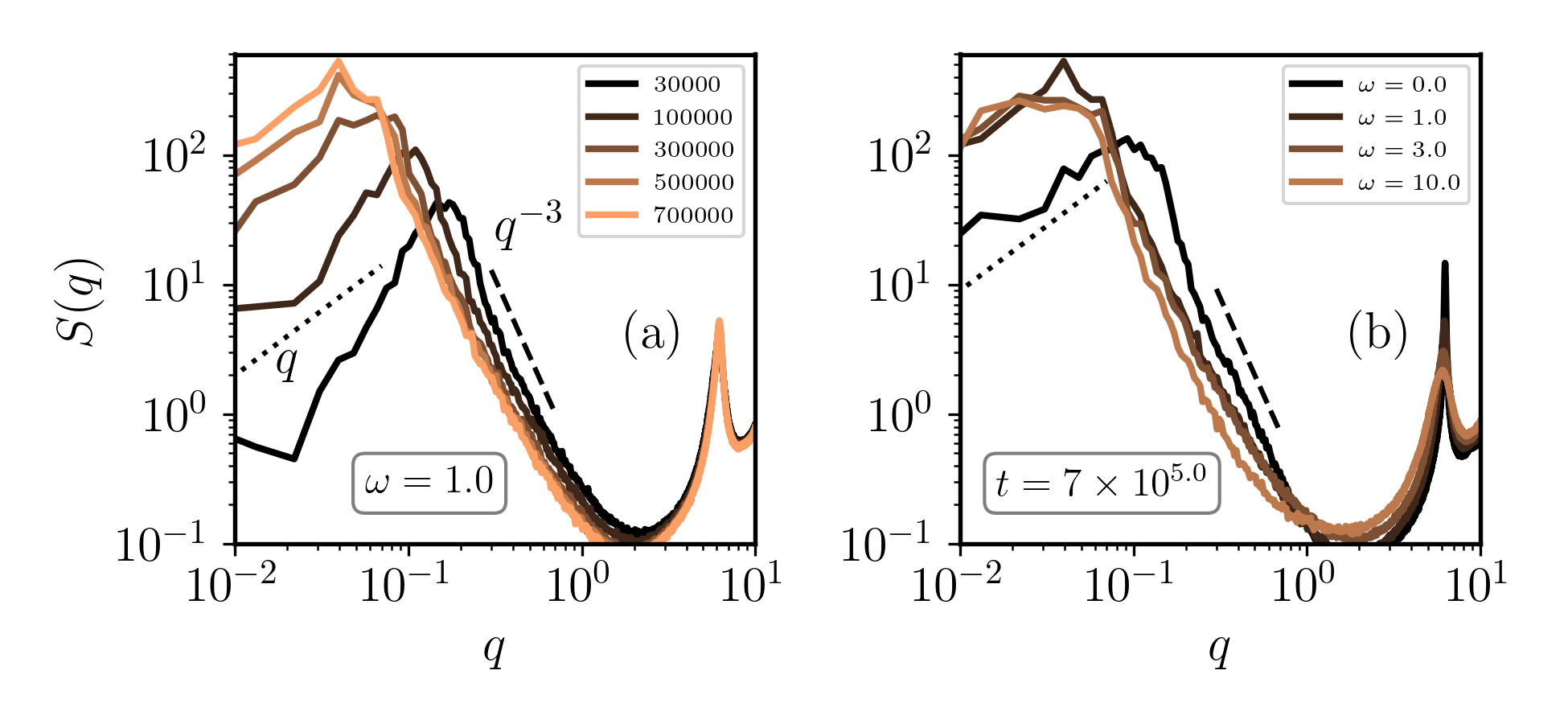}
    \includegraphics[width=0.75\linewidth]{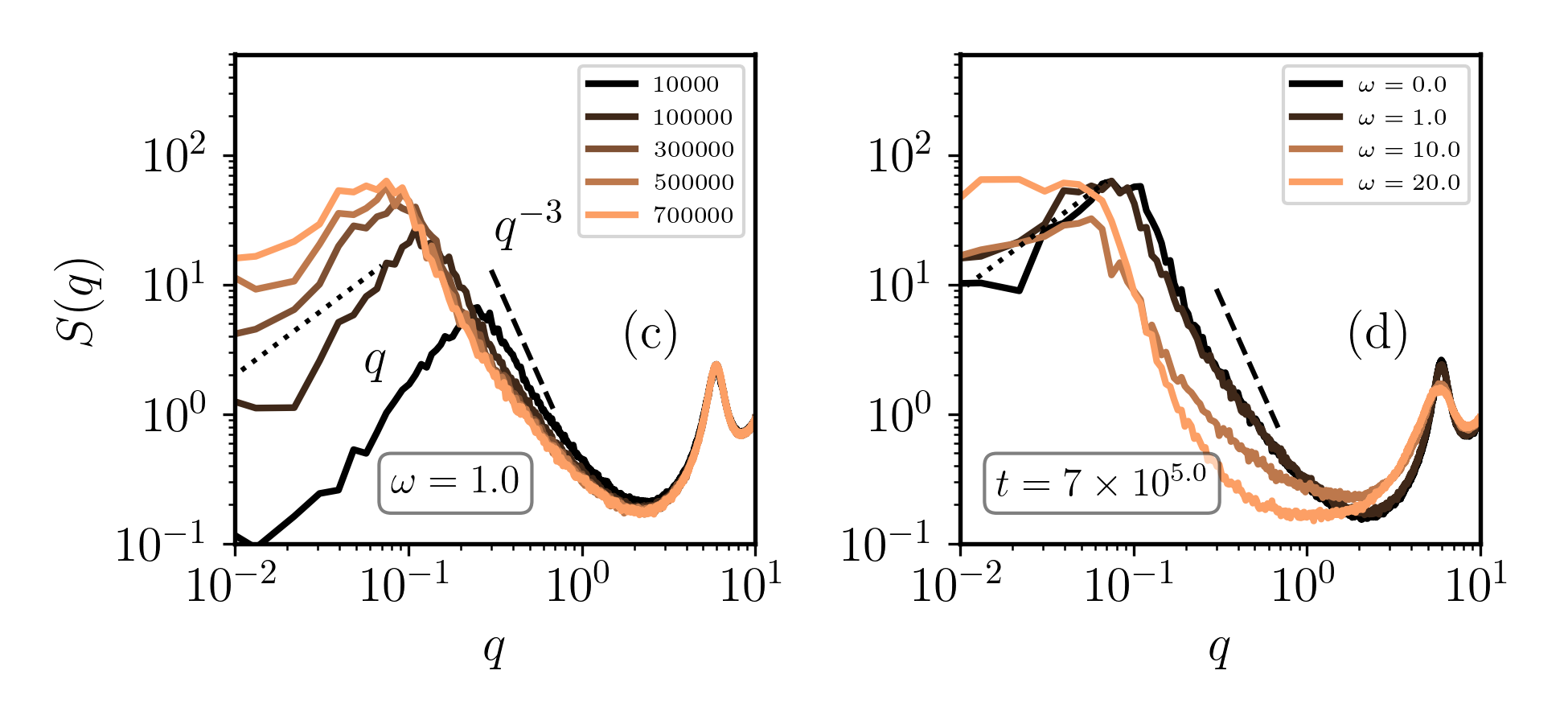}
    \caption{The structure factor $S(q, t)$ of the system, showing the coarsening process and the effect of $\omega$ on the positional order, at temperature (a-b) $k_B T=0.35$ and (c-d) $k_B T=0.47$, and fixed density $\phi = 0.40$. (a-c) The structure factor at different times for $\omega = 1.0$. (b-d) The structure factor at a fixed time for different values of $\omega$. The expected theoretical scaling for $q\rightarrow 0$ and 
    for the tail of the distributions in the scaling regime is reported with a dotted and dashed line, respectively.}
    \label{SM:structure-factor}
\end{figure}

To investigate the positional order of the particles in the dense phase and the phase separation kinetics after a quench into the coexistence region, we calculated the structure factor $S(q,t) = N^{-1}\sum_{ij} e^{i \bm{q} (\bm{r}_i(t) - \bm{r}_j(t))}$. The results are reported in Fig~\ref{SM:structure-factor}. In Fig.~\ref{SM:structure-factor}(a)-(c), we plot the spherically averaged structure factor as a function of time for a fixed chirality $\omega = 1.0$ at fixed density $\phi = 0.40$ and temperature $k_BT=0.35$, $0.47$, respectively. We find that the peak at low $q$ shifts to smaller values as time progresses, indicating that the domains of the dense phase increase in size, and the system undergoes coarsening. In Fig. Fig.~\ref{SM:structure-factor}(b)-(d), we show the structure factor at a fixed time $t=7\cdot 10^5$, which is deep inside the long-time scaling regime (the latter starts around $t \sim 10^2$ for all the values of $\omega$ considered), as a function of $\omega$, again at density $\phi = 0.40$ and temperature $k_B T=0.35$ and $0.47$. The secondary peak corresponding to the crystalline order, located at $q \approx 2\pi/\sigma_d$, becomes broader and lower as $\omega$ grows, suggesting that  $\omega$ reduces the local positional order. 

We observed that for $q \to 0$, $S(q) \sim q$ followed by a second regime $S(q) \sim q^{-d-1}$ at larger wavenumbers, 
 where $d=2$ is the dimensionality of the system, consistently with the predictions of Porod's law for systems with sharp interfaces between coexisting phases \cite{caporusso2020motility}. 

\section{Local density distributions}

\begin{figure}[!htbp]
    \centering
    \includegraphics[width=0.715\linewidth]{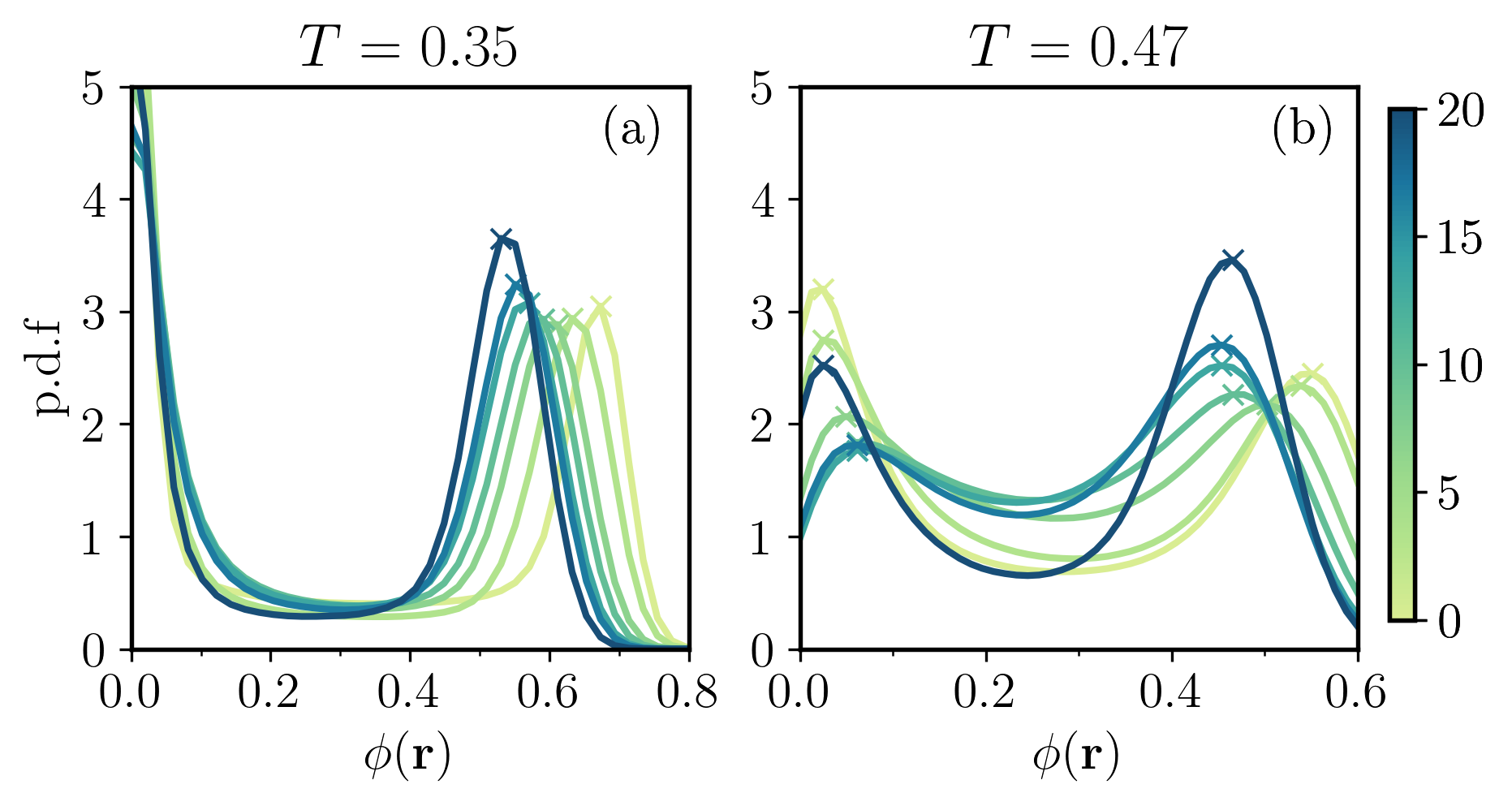}
    \caption{The probability distribution function (PDF) of the local density $\phi(\bm{r})$ for different values of  $\omega$ at two temperatures, (a) $k_B T = 0.35$ and (b) $k_B T = 0.47$. The positions of the peaks are found with a local maximum searching algorithm and are shown with crosses. The color map indicates the corresponding value of $\omega$ of each curve.}
    \label{SM:pdf}
\end{figure}

To further investigate the coexistence of chiral liquid and gas phases, we computed the probability distribution function (PDF) of the local density $\phi(\bm{r})$ in the system. The local density was obtained by dividing the simulation box into cells of linear size $5\, \sigma_d$ and counting the number of particles in each cell (although we checked that the results did not significantly depend on the chosen size of the cells). Figure \ref{SM:pdf} shows the PDF of the local density for different values of $\omega$ at the temperatures $k_BT = 0.35$ and $k_B T = 0.47$ and global density $\phi = 0.30$. The PDF has a bimodal shape, with two peaks corresponding to the liquid and gas densities. We identified the coexistence densities $\phi_{G/CL}$ by locating the positions of the peaks in the PDF. We observed that the two peaks' location does not change varying the overall global density in the coexistence region (not shown). We found that the density of the gas phase $\phi_G$ slightly increases with $\omega$, while the chiral liquid one $\phi_{CL}$ decreases slightly. This is consistent with the phase diagram shown in Fig.~1 of the main text and  Fig.~\ref{SM:phasediagram}.




\section{Stress profiles}

\begin{figure}[!htbp]
    \centering
    \includegraphics[width=0.9\linewidth]{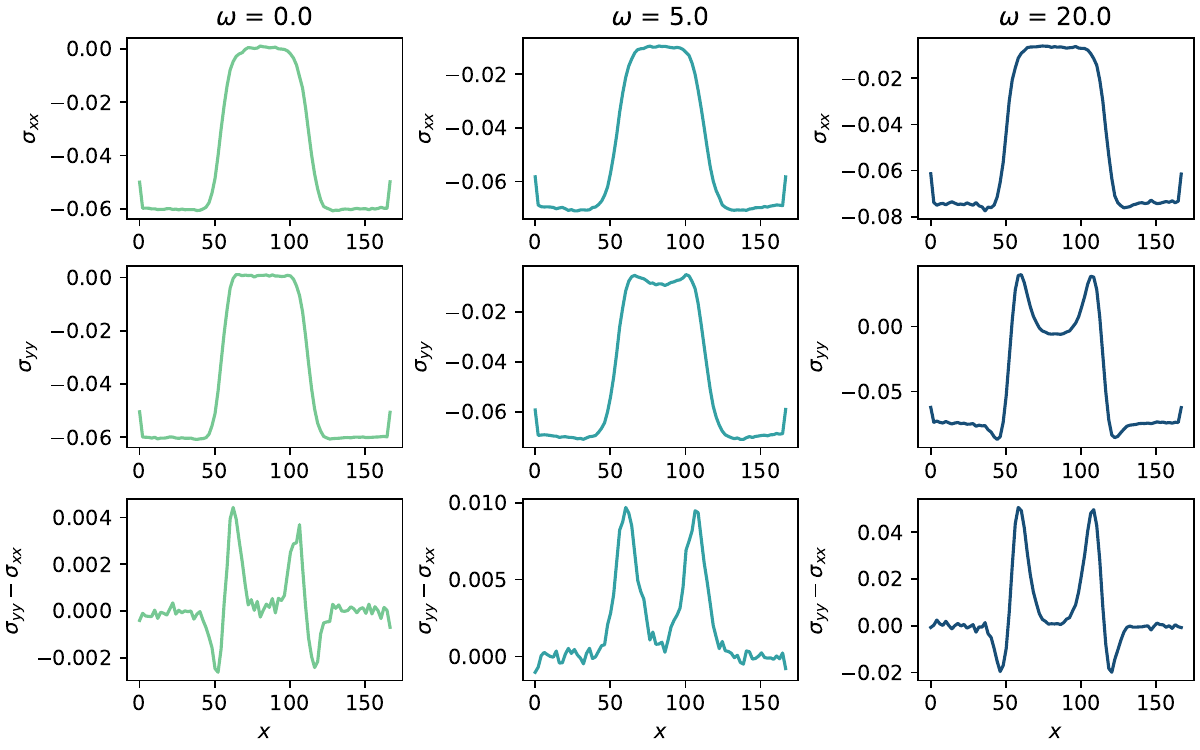}
    \caption{Profiles of the normal stress component $\sigma^{xx}$ (first row), the tangential stress component $\sigma^{yy}$ (second row), and their difference $\sigma^{yy} - \sigma^{xx}$ (third row) across a slab for  $\omega = 0$, $5$ and $20$ (respectively, first, second and third column)}.
    \label{SM:stress_comparison}
\end{figure}

We computed the profiles of the tangential stress component $\sigma^{yy}$ and the normal one $\sigma^{xx}$ in a slab configuration, oriented with the interface along the y-axis. We also computed the difference $\left(\sigma^{yy} - \sigma^{xx}\right)$, which represents the surface tension profile (see the main text for the definition). We measured the stress profile using slab configurations, where we calculated a coarse-grained version of the stress tensor over square cells of linear size $2.5\,\sigma_d$, and average it along the interface direction (y), obtaining a one-dimensional profile $\sigma^{ab}(x)$. The resulting profiles for different values of $\omega=0$, $5$ and $20$ are reported Fig.~\ref{SM:stress_comparison}.

 We notice that while the normal stress is mostly unaffected by $\omega$, the tangential stress significantly changes and is responsible for the increase in the peaks of the surface tension profile. As we mention in the main text, we hypothesize that this is caused by the currents, which are generated along the slab (y) direction and enhance collisions between particles, thus increasing the stress in the corresponding (yy) component.

\FloatBarrier


\section{Rotational viscosity as a function of density}

\begin{figure}[!htbp]
    \centering
    \includegraphics[width=0.75\linewidth]{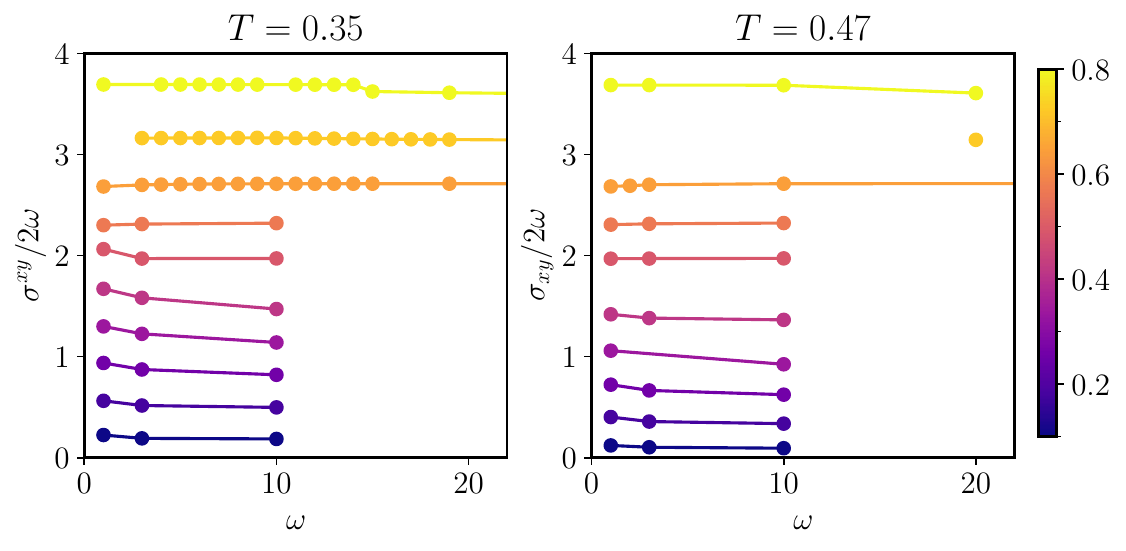}
    \caption{Rotational viscosity $\eta_R = \sigma^{xy}/(2\omega)$ for (a) $k_B T=0.35$ and (b) $k_B T=0.47$, at different values of  $\omega$, and colored according to the system's density $\phi$ (reported in the colorbar).}.
    \label{SM:stress_oemga}
\end{figure}

In this section, we report the dependence of the rotational viscosity $\eta_R = \sigma^{xy}/2\omega$  on $\omega$ and $\phi$, at temperatures $k_B T = 0.35$ and $k_B T = 0.47$, for parameter values that span all the phases of our system in Fig.~\ref{SM:stress_oemga}. As anticipated in the main text, we find that the stress tensor is largely independent of temperature and of 
 $\omega$, and mostly depends on the density of the system. 

\section{Friction coefficient}

In this section, we aim to estimate the friction coefficient $\mu$ that appears in the expression of the rotational viscosity,  $\eta_R = \mu \delta v_e / (2 \omega)$. 

\begin{figure}[!htbp]
    \centering
    \includegraphics[width=0.75\linewidth]{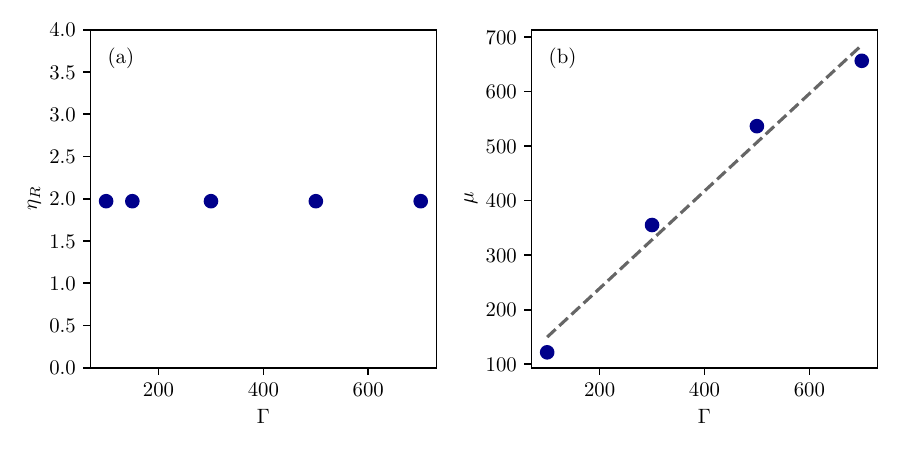}
    \caption{(a) Rotational viscosity $\eta_R$ as a function of damping coefficient $\Gamma$ in an homogeneous system at $\omega = 3.0$, $\phi = 0.60$ and $k_B T = 0.35$. (b) Friction coefficient $\mu$ as a function of the damping coefficient $\Gamma$. The dotted line is a linear fit $\eta_R = a \Gamma + b$, with $a = 0.89 \pm 0.09$ and $b = 60.69 \pm 41.16$.}
    \label{SM:mu}
\end{figure}

To this end, we first performed simulations with different values of the damping coefficient of the model $\Gamma$ in (homogeneous) systems with $\omega = 3.0$ and $\phi = 0.60$ at $k_B T = 0.35$. We computed the rotational viscosity directly from the stress tensor as $\eta_R = \sigma^{xy}/2\omega$, and found that it does not depend on $\Gamma$,  Fig.~\ref{SM:mu}(a). 

Next, we simulated a chiral fluid drop, similar to the one shown in Fig.~3 of the main text, with fixed $\omega = 3.0$ at $k_B T=0.35$ and varying $\Gamma$. We measured the edge velocity as a function of the damping coefficient $v_e(\Gamma)$ (not shown). By inverting the first expression for $\eta_R$, we obtained $\mu(\Gamma) = 2 \omega \eta_R / (\delta v_e)$, which we plotted in Fig.~\ref{SM:mu}(b) using the values of $\eta_R$ measured from the homogeneous system and $v_e(\Gamma)$ measured from the fluid drop. We fitted the data with a linear function, $\eta_R = a \Gamma + b$, with $a \sim 1$. Therefore, the data shows that $\mu \approx \Gamma$, as used in the analysis of the main text. 

\newpage
\FloatBarrier

\section{Hexatic domains and vorticity}

\begin{figure}[!htbp]
\centering
\includegraphics[width=0.9\linewidth]{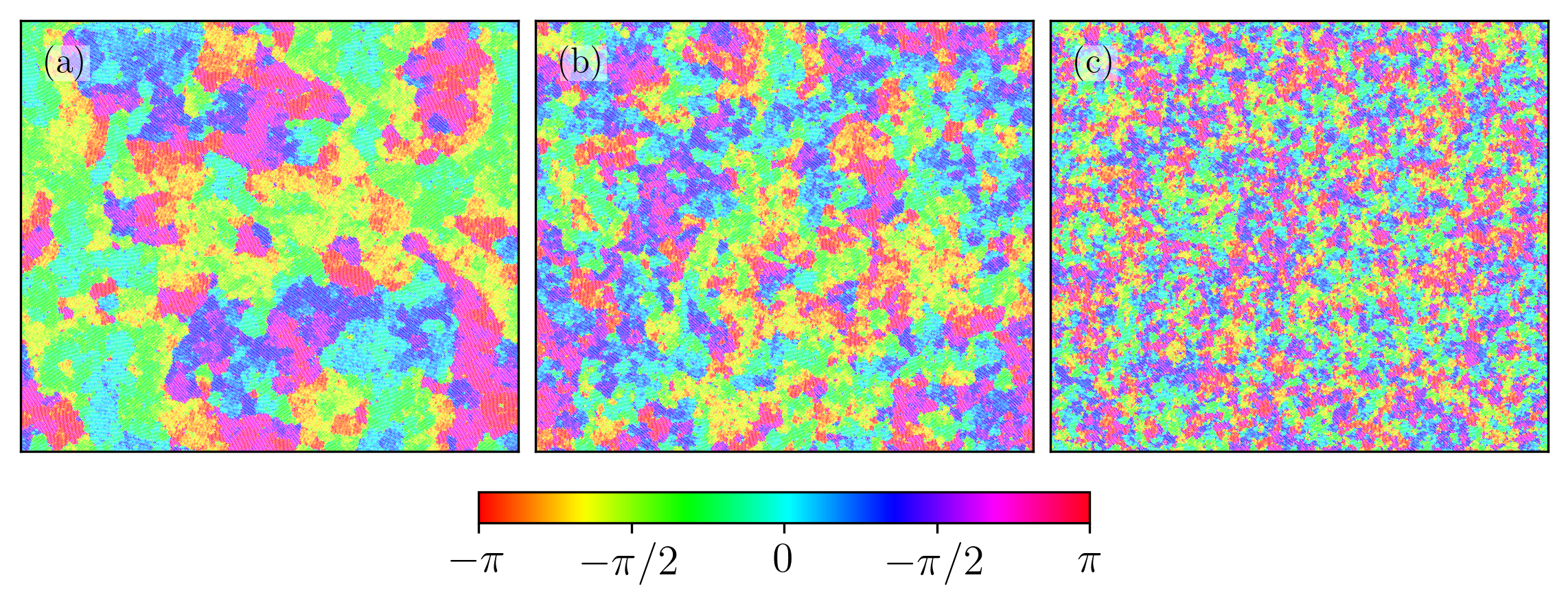}
\caption{Colormap of the hexatic local bond order parameter $\text{Arg}(\psi_6)$ (as defined in the text) for configurations at fixed $\phi = 0.70$ and different $\omega$ $\omega$: (a) $\omega = 3.0$, (b) $\omega = 5.0$ and (c) $\omega = 10.0$, respectively.}
\label{SM:hexatic_vorticity}
\end{figure}

\begin{figure}[!htbp]
    \centering
    \includegraphics[width=0.80\linewidth]{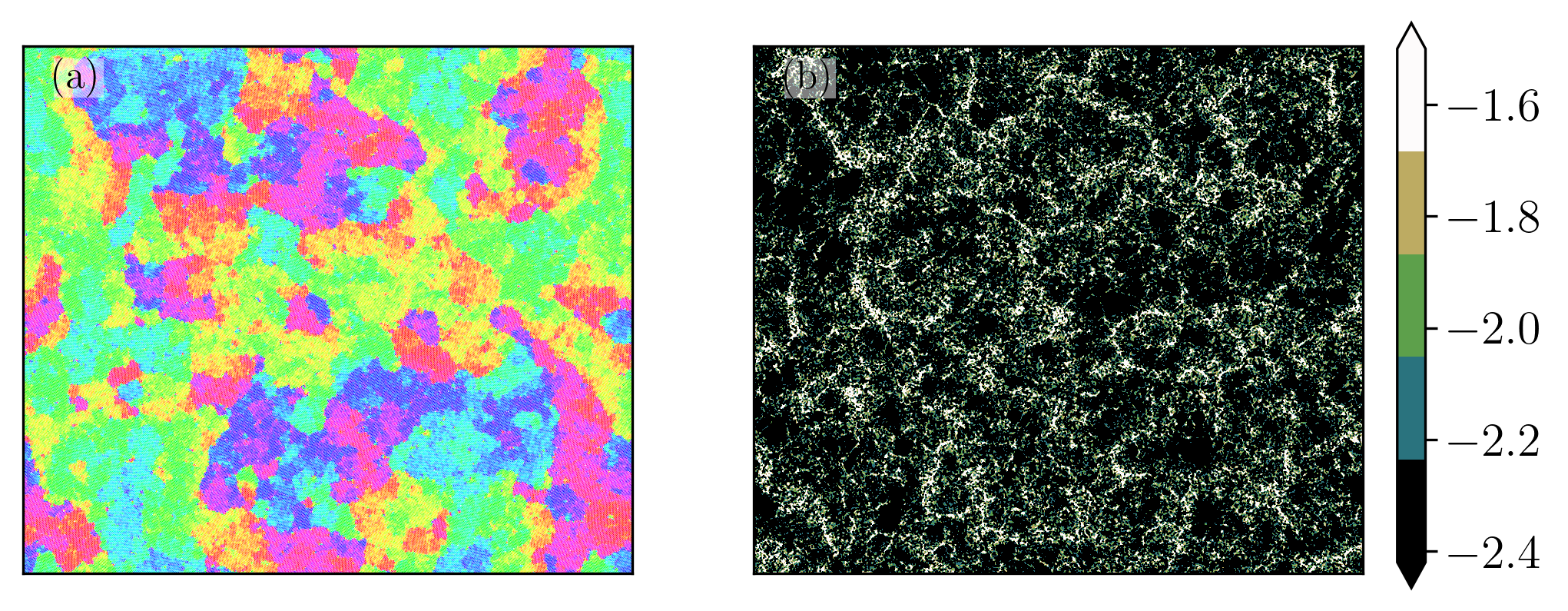}
    \caption{Conmparison between (a) the hexatic local bond order parameter with (b) the local value of the vorticity field $\Omega(\bm{r})$ for a given configuration. Each cell in (b) is colored according to the value of $\log\left[\Omega(\bm{r})/\Omega_{\text{max}}\right]$, where $\Omega_{\text{max}}$ is a normalization constant. Parameters: $k_BT=0.35$, $\phi=0.7$ and $\omega=3$.}
    \label{SM:vorticity}
\end{figure}

We report in Fig.~\ref{SM:hexatic_vorticity} some snapshots of configurations at $k_B T=0.35$ and $\phi = 0.7$ for $\omega = 3$, $5$ and $10$, that support our results in the main text that the hexatic patches, i.e, domains with the same local bond order parameter $\text{Arg}(\psi_6)$, become smaller upon increasing  $\omega$. We also compare the local bond parameter map of the configuration at $\omega = 3.0$ with its vorticity field $\bm{\Omega} = \bm{\nabla} \times \bm{v}$, Fig.~\ref{SM:vorticity}. We observe the largest vorticity regions are localized at the grain boundaries between hexatic patches, with the bulk having a lower value.  

\bibliography{chiral_abbrv}